\documentclass[10pt, journal, twocolumn]{IEEEtran}

\usepackage{times}
\usepackage{relsize}

\usepackage{amsmath}  
\usepackage{amssymb}  
\usepackage{mathrsfs} 

\usepackage{theorem}  
\usepackage{cite}     
\usepackage{comment}  

\usepackage{upref}
\usepackage{amsfonts}

\usepackage{verbatim}

\usepackage[dvipsnames,usenames]{color}

\usepackage{graphicx}
\usepackage{subfigure}
\usepackage{psfrag}
\usepackage{algpseudocode}
\usepackage[export]{adjustbox}
\usepackage{arydshln}

\newcommand{\be}[1]{\begin{equation}\label{#1}}
\newcommand{\ee}{\end{equation}}

\newcommand{\bc}{\begin{center}}
\newcommand{\ec}{\end{center}}

\newcommand{\cF}{{\cal F}}
\newcommand{\cG}{{\cal G}}

\newcommand{\cR}{{\cal R}}

\newcommand{\cX}{{\cal X}}
\newcommand{\cY}{{\cal Y}}

\renewcommand{\leq}{\leqslant}

\renewcommand{\geq}{\geqslant}

\newcommand{\Cref}[1]{Co\-rol\-la\-ry\,\ref{#1}}

\theoremstyle{plain} \theorembodyfont{\normalfont\slshape}

\newtheorem{thm}{Theorem$\!$}
\newenvironment{theorem}{\begin{thm}\hspace*{-1ex}{\bf.}}{\end{thm}}

\newtheorem{prop}{Proposition$\!$}

\newtheorem{lem}{Lemma$\!$}

\newtheorem{conj}[thm]{Conjecture$\!$}

\newtheorem{cor}{Corollary$\!$}

\newtheorem{cl}{Claim$\!$}

\newtheorem{defi}{Definition$\!$}
\newenvironment{definition}{\begin{defi}\hspace*{-1ex}{\bf .}}{\end{defi}}

\newtheorem{const}{Construction$\!$}

\newtheorem{algr}{Algorithm$\!$}

\theorembodyfont{\normalfont}

\newtheorem{exam}{Example$\!$}

\newtheorem{remrk}{Remark$\!$}
\newenvironment{remark}{\begin{remrk}\hspace*{-1ex}{\bf .}}{\end{remrk}}

\newlength{\paragraphindent}
\newlength{\widthone}
\newlength{\widthtwo}
\newlength{\widththree}
\newlength{\colwidthtemp}
\definecolor{Codecolor}{named}{White}  

\newcommand{\Copen}{\mbox{\{\kern-5.50pt\{}}
\newcommand{\Cclose}{\mbox{\}\kern-5.50pt\}}}
\newcommand{\Cslash}{\mbox{$\backslash\kern-6.02pt\backslash$}}

\begin{document}

\title{$\,$
	{ \textbf{ Robustness of Neural Networks against Storage Media Errors}}}
\author{\large Minghai~Qin,  Chao~Sun, and Dejan Vucinic \\
	Western Digital Research, San Jose, CA, USA,\\
	\{Minghai.Qin, Chao.Sun, Dejan.Vucinic\}@wdc.com
	\vspace{-1em}
	}

\maketitle

\begin{abstract}
We study the trade-offs between storage/bandwidth and prediction accuracy of neural networks that are stored in noisy media. Conventionally, it is assumed that all parameters (e.g., weight and biases) of a trained neural network are stored as binary arrays and  are error-free. This assumption is based upon the implementation of error correction codes (ECCs) that correct potential bit flips in storage media. However, ECCs add storage overhead and cause bandwidth reduction when loading the trained parameters during the inference. We study the robustness of deep neural networks when bit errors exist but ECCs are turned off for different neural network models and datasets. It is observed that more sophisticated models and datasets are more vulnerable to errors in their trained parameters. We propose a simple detection approach that can universally improve the robustness, which in some cases can be improved by orders of magnitude. We also propose an alternative binary representation of the parameters such that the distortion brought by bit flips is reduced and even theoretically vanishing when the number of bits to represent a parameter increases.
\end{abstract}

\section{Introduction} \label{sec:intro}

Neural networks (NNs)~\cite{Schmidhuber15,HTF01} are layered networks that try to fit the function of neurons in a human brain during object recognition, decision making, etc. They are one of the most widely-used machine learning techniques due to their good performance in practice. Some variants of neural networks are shown to be more suitable for different learning applications. For example, deep convolutional neural networks (CNNs)~\cite{JKR09,KSH12} are found to be effective in recognizing and classifying images. Recurrent neural networks (RNNs)~\cite{GLF09,HAF14} provides stronger performance in sequence prediction, e.g., speech or text recognition. Compared to standard feed-forward neural networks with full connections, CNNs have much less number of connections in the convolutional layers and thus much fewer parameters to train, possibly avoiding the over-fitting problems. RNNs have memory units like long-short-term-memory (LSTM)~\cite{GSKSS16} that can be trained without vanishing/exploding gradient problems. The neural networks are determined by the connections between neurons, each of which is associated with a trainable parameter called a {\em weight}. There is another parameter associated with each neuron called a {\em biase}. Since a bias can be viewed as a weight from a neuron with constant input, we will indiscriminately call it a weight as well. The set of all trainable weights are trained by back-propagation~\cite{RB93,Nielsen89} in the training process.

In order to fit highly non-linear functions and thus achieve a high rate of correctness in practice, neural networks usually contain millions to billions of weights trained from a large dataset in a careful manner to avoid over-fitting problem. In the current technology, the weights are usually stored in non-volatile memories (NVMs) and are loaded to CPU/GPU caches to make predictions for input (e.g., images, voices, or texts) during inference phase. NVMs are noisy media where bit errors can happen during writing, reading, or retention. Error correction codes (ECCs)~\cite{RL09} are ubiquitously used in NVM systems to guarantee data reliability by adding 10\% to 20\% storage redundancy.  There are two major reasons why we study the robustness of neural networks when weights stored in noisy NVM media are not fully recovered by ECCs. Firstly, the GPU caches have limited size (usually in Giga-byte range) but the size of the neural network models grow fast since the cost of gathering big data becomes smaller. If caches in a single or multiple GPUs cannot hold all weights of a neural network model, the bandwidth of loading them from NVMs to GPU caches would affect the system performance, especially for high throughput applications, such as video recognition during self-driving where the number of frames processed per second positively correlates to the safety factor. The storage overhead brought by ECCs will add latency and reduce the throughput of the NVM chips. Secondly, there is a growing number of research on using NVMs themselves (or built-in unit on NVM chips) for calculations in neural networks instead of CPU/GPU centered calculation. The benefit of this in-memory computing comes from the highly parallelism of NVM systems, but ECCs might have to be weakened or abandoned depending on the design of the computing system.


Robustness of neural networks has been studied against random and adversarial noise.~\cite{CW17} provides adversarial attack algorithms on input and defensive distillation towards evaluating the robustness of neural networks.~\cite{ZSLG16} proposed training algorithms that address the issue of output instability when the input is slightly distorted.~\cite{CBD15} and~\cite{MAAEM16} studied neural networks with binary or ternary weights, whose training algorithms are adjusted.  

In this paper, we study the robustness of a trained neural networks when they are stored in noisy media. The neural networks are trained using floating-point number calculations on GPUs and the real-valued weights are stored in binary arrays. We will review some mappings from real numbers to fixed-length binary arrays. Each bit in the binary array will be flipped identically and independently with some probability, called {\em raw bit error rate} (RBER), and the accuracy of the distorted neural networks will be examined. We then propose a simple operation to improve the accuracy by adding a check bit for each weight. Several datasets and neural network models are examined to confirm the universal benefits brought by the proposal. Finally, we provide some theoretical analysis on the binary representations of real numbers such that the bit-flip-incurred distortion can be reduced.

\section{Preliminaries}
\subsection{Neural networks and notations}
A neural network contains input neurons, hidden neurons, and output neurons. It can be viewed as a function $f:\cX \rightarrow \cY$ where the input $x\in\cX\subseteq\cR^n$ is an $n$-dimensional vector and the output $y\in\cY\subseteq\cR^m$ is an $m$-dimensional vector. In this paper, we focus on classification problems where the output $y=(y_1,\ldots,y_m)$ is usually normalized such that $\sum_{i=1}^m y_i = 1$ and $y_i$ can be viewed as the probability for some input $x$ to be categorized as the $i$-th class. The normalization is often done by the softmax function that maps an arbitrary $m$-dimensional vector $\hat y$ into normalized $y$, denoted by $y=softmax(\hat y)$, as $y_i = \frac{\exp(\hat y_i)}{\sum_{i=1}^m \exp(\hat y_i)}, i = 1,\ldots,m$. For top-$k$ decision problems, we return the top $k$ categories with the largest output $y_i$. In particular for hard decision problems where $k=1$, the classification results is then $\arg\max_i y_i, i=1,\ldots,m$.

A feedforward neural network $f$ that contains $n$ layers (excluding the softmax output layer) can be expressed as a concatenation of $n$ functions $f_i, i = 1,2,\ldots,n$ such that $f = f_n(f_{n-1}(\cdots f_1(x)\cdots))$. The $i$th layer $f_i: \cX_i\rightarrow \cY_i$ satisfies $\cY_i\subseteq \cX_{i+1}$, $\cX_1 = \cX$. The output of last layer $\cY_n$ is then fed into the softmax function. The function $f_i$ is usually defined as
\[
f_i(x) = \sigma(W \cdot x + b),
\]
where $W$ is the weights matrix, $b$ is the bias vector, and $\sigma$ is an element-wise activation function that is usually nonlinear, e.g., sigmoid and rectified linear unit (ReLU). Both $W$ and $b$ are trainable parameters.

\begin{figure}[htbp]
\centering
\includegraphics[width=1\linewidth]{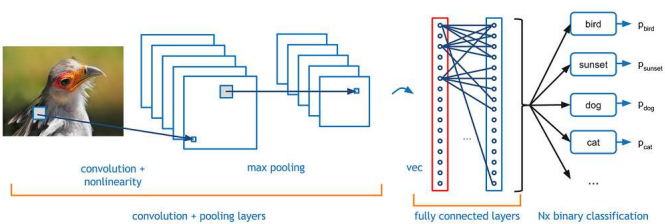}
\caption{Example of a CNN}
\label{fig:cnn_example}
\end{figure}

A convolutional neural network (CNN) (Fig.~\ref{fig:cnn_example}) is a special class of feedforward neural network that has local weights constraints, e.g., the weights connecting neurons are all zeros except for a few pair of neurons between adjacent layers, and the value of weights between different pair of neurons in two layers with similar spatial relationships are forced to be the same. Therefore, a CNN layer has much less parameters to train compared to a fully connected layer and is good at extract local features from the previous layer, which enable it to be the state-of-art technique for image recognition problems. 

\begin{figure}[htbp]
\centering
\includegraphics[width=\linewidth]{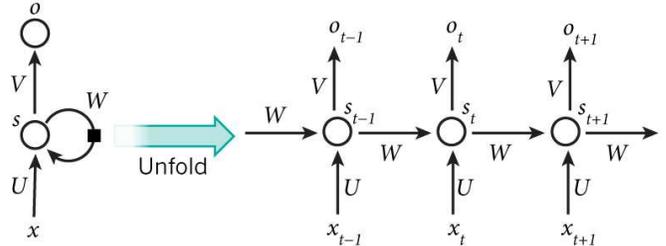}
\caption{Unfolding a RNN}
\label{fig:rnn}
\end{figure}

A recurrent neural network (RNN) is a special class of neural networks that has directed cycles, which enable it to create internal states and exhibit temporal behaviors. A RNN can be unfolded (Fig.~\ref{fig:rnn}) in time to form a feedforward neural network for training purposes. One of the most widely used neurons to store the states is LSTM, consisting of forget-gate, update-gate, and output-gate. Back-propagation algorithms can be applied from the last output neurons backwards to train all weights in the RNN.

In this paper, we focus on exploiting the robustness of deep CNNs and RNNs on different datasets.

\subsection{Real-valued weights and their binary representations}
Assume we have a trained neural network whose weights are all real numbers. In order to store the real-valued weights into storage media, each of them needs to be converted into binary arrays of certain length. Several ways to map real numbers to binary arrays exist, among which the most widely used is IEEE Standard for Floating-Point Arithmetic (IEEE 754).  In this paper, we focus on 16-bit representations of real numbers since the precision of 16 bits is high enough to avoid degradation of inference performance due to quantization errors. 

\subsubsection{IEEE Standard for Floating-Point Arithmetic}
IEEE 754 provides guidelines to represent a real number $r$ as $(-1)^s \times c \time 2^q$ by 16, 32, 64, 128, and 256 bits, where $s$ is 1-bit of sign, $c$ is a significand, and $q$ is an exponent. For 16-bit representation, 5 bits are assigned to the exponent with bias equal 15 and the rest 11 bits are used for the 10-bit significand and the 1-bit sign. The largest number that can be represented is $(2-2^{-10})\times2^{15} = 65504$ which is much larger than the weights in any neural network models.
Since almost all weights are between $-1$ and $+1$ (or even much closer to 0), the representation precision is at least $2^{-11}$ (numbers between $0.5$ and $1$), and the precision will increase exponentially when the real number decreases exponentially to 0. 

With the presence of media errors, IEEE standard for floating-point arithmetic is not a proper representation of real-valued weights. The major weakness is due to the large range the IEEE 754 standard can represent. In particular, if the most significant bit in the exponent is erroneous, the value of that weight can be inadvertently set to a very large value. For example, the binary string $0\ 01101\ 0101010101$ represents $(-1)^0 \times 2^{13-15} \times 1.3330078125 \approx 0.33$, but if the second bit is flipped and the string becomes $0\ 11101\ 0101010101$, it will represent $(-1)^0 \times 2^{29-15} \times 1.3330078125 = 21840$. This large value of weight will totally destroy the learned neural network. Since the number of weights is large and each weight has a few ``vulnerable'' bits (e.g., some most significant bits in the exponents), there is rarely any robustness in the trained neural networks against the media errors.

\subsubsection{Binary expansion}\label{sec:bin_exp}
In order to avoid the large distortion incurred by IEEE 754 standard floating-point number representation, a direct quantization of real numbers between the minimum-valued weight and the maximum-valued weight can be implemented. Assume the minimum and maximum weight is denoted by $w_{\textrm{min}}$ and $w_{\textrm{max}}$, respectively. To convert a weight into a binary array, the interval $[w_{\textrm{min}}, w_{\textrm{max}}]$ can be quantized into $2^q$ consecutive subintervals with boundaries 
$$w_{\textrm{min}}=b_0<b_1<\cdots<b_{2^q-1}=w_{\textrm{max}},$$
 where $q$ is the number of bits to represent a weight. For all weights $w\in[w_{\textrm{min}}, w_{\textrm{max}}]$, if $w$ is in the $i$th interval, i.e., $b_i\leq w < b_{i+1}$, then $w$ is represented by the $q$-bit binary expansion of the decimal integer $i$ as $(b_0,b_1,\ldots,b_{q-1})$. When the weights are loaded from binary storage array for calculation, it can be recovered by
 \[
 w_{\textrm{rec}} = w_{\textrm{min}} + \frac{w_{\textrm{max}}-w_{\textrm{min}}}{2^q} \times \left( \frac{1}{2} + \sum_{j=0}^{q-1}b_j2^{j}\right).
 \]
 
 The binary expansion representation of weights avoids exploding problem in IEEE 754 standard, but it also has weakness in that all bits have different robustness against errors. If the most significant bit is in error, then the distortion is as large as half of the range, i.e., $|w_{\textrm{erroneous}}-w_{\textrm{true}}|=\frac{1}{2}(w_{\textrm{max}}-w_{\textrm{min}})$; on the other hand, if the bits in the non-significant bits are in error, the distortion is too small to affect the performance. The unequal vulnerability of bits might increase the variance and cause unpredictability and instability of neural networks. In Section~\ref{sec:HDB}, a Hamming-distance-based binary representation of real numbers will be proposed to reduce the unequal vulnerability of the $q$ bits.

\section{Robustness of Neural Networks}
In this section, we provide a general view of robustness of neural networks against media errors, where a weight is represented by the binary expansion of the index of its located intervals. First, we train a neural network for a specific task, e.g, handwritten digit recognition, using floating-point calculation by GPU. We use 16 bits to represent each trained weight and it can be seen from our experiments that quantization error is negligible with this precision. Suppose the total number of weights is $N$, then the total number of bits to represent the neural network (excluding the architecture) is $16N$. We will model the media errors as passing through a binary symmetric channel (BSC)~\cite{CT91}, where each bit is flipped with probability $p$. The value of $p$ is called the raw bit-error rate (RBER) of the media. Therefore, the average total number of bit errors is $16Np$. One {\em distorted model} is then obtained by using this distorted set of weights. For each $p$, multiple distorted models are saved and then used to evaluate the inference performance and approximate the statistics.  Note that the robustness of the neural network models is the major concern, thus we did not try to train and achieve the best accuracy for undistorted models, but rather explore the performance loss between the undistorted and distorted model. We will later provide schemes to reduce the performance loss for the same RBER. 

\subsection{MNIST}
The MNIST database is a large database of handwritten digits that is commonly used for image processing and machine learning. It includes 60000 training images and 10000 test images of size $28 \times 28$. We train a convolutional neural network to classify the 10 digits with the architecture in Table~\ref{tab:mnist}. According to the table, the convolutional layers have filters of size $3\times 3$ and we use ``valid'' as the padding scheme. The max pooling layers have filters of size $2\times 2$. The output of the last fully connected layer is passed through a softmax layer to make a prediction. The prediction accuracy for this trained model is $0.9961$ when there is no media error.
\begin{table}[htbp]
	\center
	\caption{CNN architecture for MNIST}
	\label{tab:mnist}
\begin{tabular}{|c:c:c|}
	\hline Layer & Output shape & No. of trainable parameters \\ 
	\hdashline Conv. 2D(3,3) & $(26,26,32)$ & 320 \\ 
	\hdashline Conv. 2D(3,3) & $(24,24,32)$ & 9248 \\ 
	\hdashline Maxpooling & $(12,12,32)$ & 0 \\ 
	\hdashline Conv. 2D(3,3) & $(10,10,64)$ & 18496 \\ 
	\hdashline Conv. 2D(3,3) & $(8,8,64)$ & 36928 \\ 
	\hdashline Maxpooling & $(4,4,64)$ & 0 \\ 
	\hdashline Fully connected & $(256)$ & 262400 \\ 
	\hdashline Fully connected & $(10)$ & 2570 \\ 
	\hline 
\end{tabular} 
\end{table}

Table~\ref{tab:mnist_accu} shows the average prediction accuracy for different raw bit-error-rate of the media, where all weights are represented by the binary expansion. Note that the accuracy of RBER$=0$ is 0.0002 less than the undistorted model, which means the loss of the quantization from 32-bit floating point representation to 16-bit binary expansion is 0.0002. Each data is obtained by averaging the accuracy of 40 distorted models sampled independently.
\begin{table}[htbp]
	\center
	\caption{Prediction accuracy for MNIST}
	\label{tab:mnist_accu}
\begin{tabular}{|c:c:c:c:c:c|}
	\hline  RBER & 0 & 0.001 & 0.002 & 0.003 & 0.004  \\ 
    \hdashline Ave. Acc. & 0.9959 & 0.9952 & 0.9945 & 0.9937 & 0.9924  \\ 
	\hline  RBER & 0.005 & 0.006 & 0.007 & 0.008 & 0.009  \\ 
    \hdashline Ave. Acc. & 0.9906 & 0.9885 & 0.9847 & 0.9816 & 0.9754 \\ 
	\hline  RBER & 0.01 & 0.011 & 0.012 & 0.013 & 0.014   \\ 
	\hdashline  Ave. Acc. &  0.96952 & 0.9576 & 0.94875 & 0.9339 & 0.9224   \\ 
	\hline 
\end{tabular} 
\end{table}

\begin{figure}[htbp]
\centering
\includegraphics[width=1\linewidth]{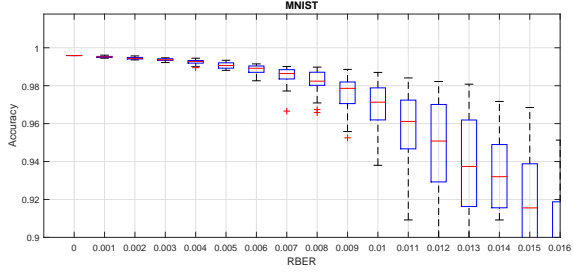}
\caption{Box plot of MNIST database with different probability of media errors}
\label{fig:mnist_naive_box}
\end{figure}

Fig.~\ref{fig:mnist_naive_box} shows the box plot of the 40 distorted models for each RBER. The top and bottom of the box are the 25th and 75th percentile of the sample, where the height of the box is called the {\em interquartile range}. The line in the box is the sample median. The lines extending the top and bottom of the box is called {\em  whiskers} which is by default 1.5 interquartile range away from the top and bottom of the box, respectively. The observations beyond the whiskers are outliers.
 It can be seen that the performance variance increases dramatically with the increment of RBER.

In order the measure the robustness of neural networks against media errors, we introduce a robustness measure $R(x)$, which is defined as follows. Suppose the undistorted model has prediction accuracy $A\in[0,1]$, then $R(x),x\in[0,1]$ is defined as the maximum RBER that the average accuracy is larger than or equal to $Ax$. In the sequel, we set $x=0.95$ as default, that is, $R=R(0.95)$ is the maximum RBER that the model can tolerate when the prediction accuracy is at least $95\%$ of the undistorted model. It can be seen that the robustness of the CNN for the MNIST is $R\approx 0.012$ ($0.012$ is the largest RBER with accuracy greater than $0.9961*0.95=0.9463$).

\begin{figure}[htbp]
\centering
\includegraphics[width=1\linewidth]{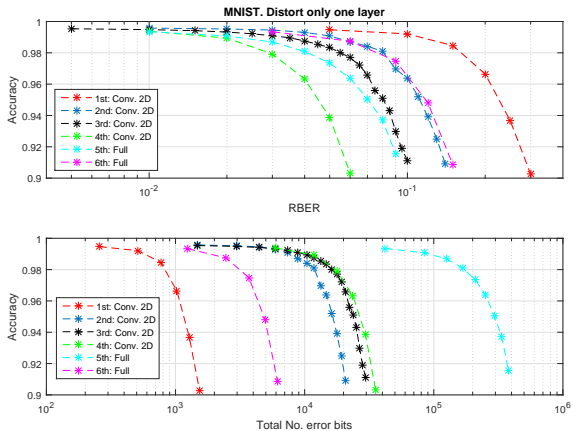}
\caption{Prediction accuracy when only one of the six layers has bit errors.}
\label{fig:mnist_one_layer}
\end{figure}

In order to explore the robustness of each convolutional layer or fully connected layer against errors, we test the prediction accuracy when weights in each layer are distorted individually. The top figure in Fig.~\ref{fig:mnist_one_layer} shows the average prediction accuracy over 40 distorted models for each RBER where bit errors appear in only one of the six layers. It can be observed that the robustness is decreasing from the first to the fourth convolutional layer, possibly due to the fact that the number of weights is largely increased from the first to the fourth.
In order to take the total number of weights into account, the bottom figure in Fig.~\ref{fig:mnist_one_layer} shows the average prediction accuracy versus the total number of bit errors in each layer. It is reasonable to see that layers with more weights can tolerate more bit errors at the same prediction accuracy.

\subsection{CIFAR-10}
The CIFAR-10 dataset consists of 60000 $32\times 32$ color images in 10 classes, with 6000 images per class. It is partitioned into 50000 training images and 10000 test images. The 10 classes include airplane, automobile, bird, cat, deer, dog, frog, horse, ship, and truck, where all classes are mutually inclusive. 

VGG nets~\cite{SZ14} are convolutional neural networks that have been shown to achieve a good performance for image recognition. The input of VGG-16 and VGG-19 are both $224\times 224$, which is larger than CIFAR-10 images, thus we skip the last few layers including two pooling layers to obtain a shortened VGG-like architecture summarized in Table~\ref{tab:cifar10}.

\begin{table}[htbp]
	\center
	\caption{CNN architecture for CIFAR-10}
	\label{tab:cifar10}
\begin{tabular}{|c:c:c|}
	\hline Layer & Output shape & No. of trainable parameters \\ 
	\hdashline Conv. 2D & (32,32,64) & 1792 \\ 
	\hdashline Conv. 2D & (32,32,64) & 36928 \\ 
	\hdashline Max pooling & (16,16,64) & 0 \\ 
	\hdashline Conv. 2D & (16,16,128) & 73856 \\ 
	\hdashline Conv. 2D & (16,16,128) & 147584 \\ 
	\hdashline Max pooling & (8,8,128) & 0 \\ 
	\hdashline Conv. 2D & (8,8,256) & 295168 \\ 
	\hdashline Conv. 2D & (8,8,256) & 590080 \\ 
	\hdashline Conv. 2D & (8,8,256) & 590080 \\ 
	\hdashline Max pooling & (4,4,256) & 0 \\ 
	\hdashline Fully connected & (256) & 1048832 \\ 
	\hdashline Fully connected & (10) & 2570 \\ 
	\hline 
\end{tabular} 
\end{table}

\begin{figure}[htbp]
\centering
\includegraphics[width=1\linewidth]{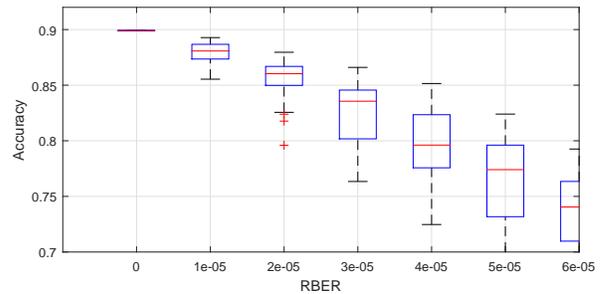}
\caption{Box plot of CIFAR-10 for different RBERs}
\label{fig:cifar10_box}
\end{figure}

Table~\ref{tab:cifar10_accu} and Fig.~\ref{fig:cifar10_box} show the accuracy and the box plots for each RBER for the CIFAR-10 dataset using the CNN architecture in Table~\ref{tab:cifar10}. The robustness of the CNN for CIFAR-10 is therefore $R(0.95)\approx 2\times 10^{-5}$, much less than the robustness of the CNN for MNIST dataset.
\begin{table}[htbp]
	\center
	\caption{Prediction accuracy for CIFAR-10}
	\label{tab:cifar10_accu}
\begin{tabular}{|c:c:c:c:c:c|}
	\hline RBER & 0 & 1e-5 & 2e-5 & 3e-5 & 4e-5 \\ 
	\hdashline Ave. Accu. & 0.8991 & 0.8789 & 0.8555 & 0.8265 & 0.7984 \\ 
	\hline 
\end{tabular} 
\end{table}

\begin{figure}[htbp]
\centering
\includegraphics[width=1\linewidth]{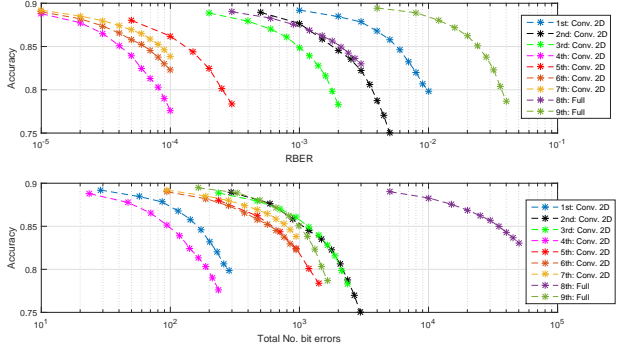}
\caption{Prediction accuracy when only one of the nine layers has bit errors.}
\label{fig:cifar10_one_layer}
\end{figure}

Fig.~\ref{fig:cifar10_one_layer} shows the accuracy versus RBER and versus total number of bit errors when only one of the nine layers includes weights with errors. Similar to the CNN for MNIST, the robustness is increasing as from the first convolutional layer with the minimum weights to the later layers with more weights.

\subsection{IMDB}
IMDB is one of the  most popular and authoritative source for movie, TV and celebrity content. The 50000 movie reviews are equally partitioned into training set and test set. Each review is categorized as either ``positive'' or ``negative''. 

Recurrent neural networks (RNNs) enables to remember ``states'', which are proved to be useful in understanding languages. We use a simple RNN layer consisting of 100 LSTM units after an embedding layer and a one-dimensional convolutional layer. The embedding layers transform the  word vector into its corresponding word embeddings~\cite{MSCCD13}. The detailed architecture and the number of trainable parameters in the model the listed in Table~\ref{tab:imdb}.

\begin{table}
\centering
\caption{RNN architecture for IMDB}
\label{tab:imdb}
\begin{tabular}{|c:c:c|}
	\hline Layer & Output shape & No. of trainable parameters \\ 
	\hdashline Embedding & (500,32) & 160000 \\ 
	\hdashline Conv. 1D & (498,32) & 3104 \\ 
	\hdashline Max pooling & (124,32) & 0 \\ 
	\hdashline LSTM & (100) & 53200 \\ 
	\hdashline Full Connect. & (1) & 101 \\ 
	\hline 
\end{tabular} 
\end{table}

\begin{figure}[htbp]
\centering
\includegraphics[width=1\linewidth]{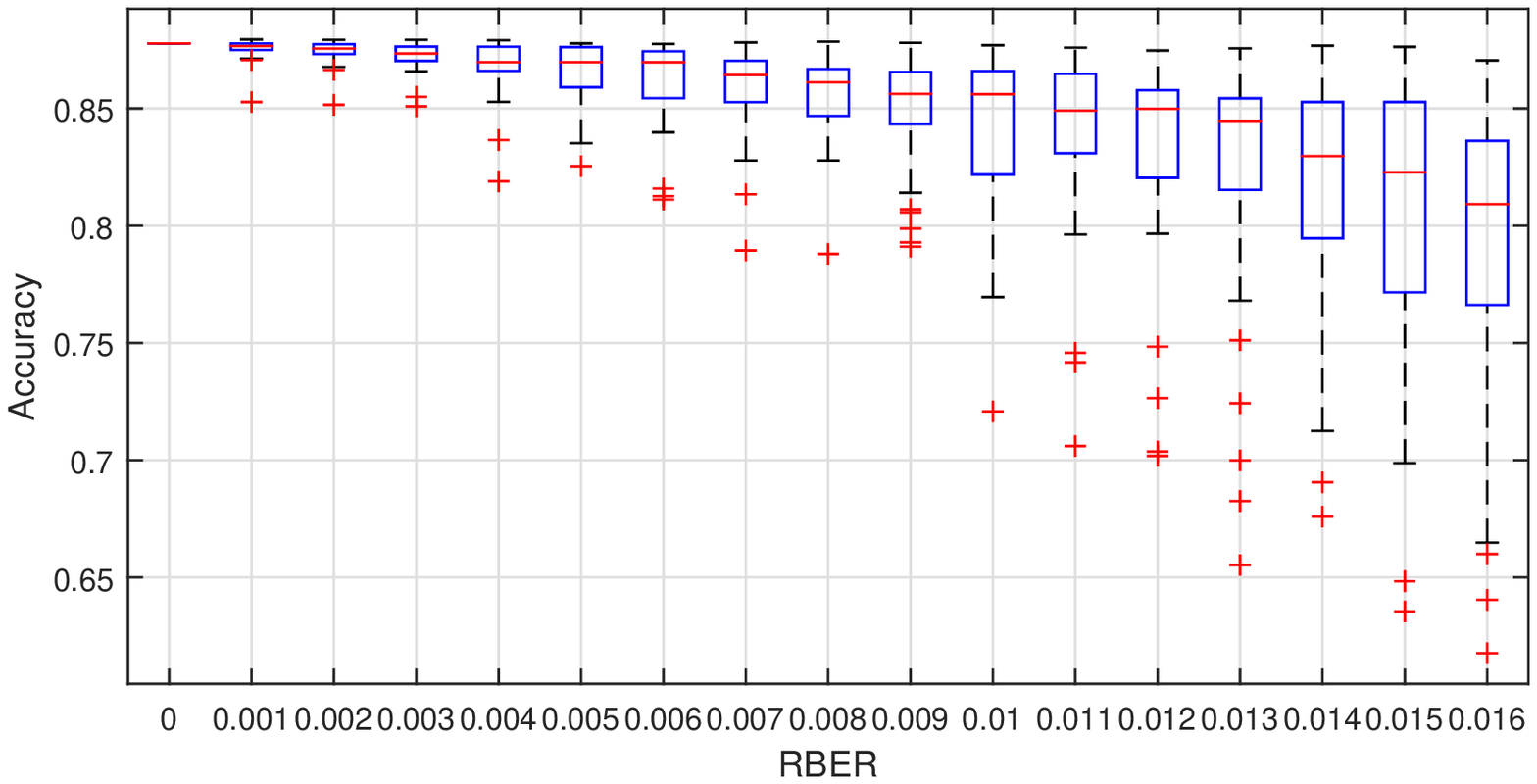}
\caption{Box plot of accuracy for IMDB when the LSTM layer has bit errors}
\label{fig:imdb_box_lstm}
\end{figure}

\begin{figure}[htbp]
\centering
\includegraphics[width=1\linewidth]{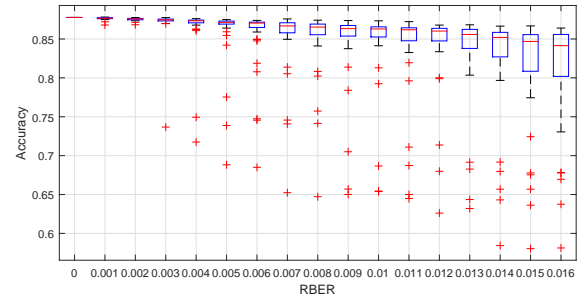}
\caption{Box plot of accuracy for IMDB when the embedding layer has bit errors}
\label{fig:imdb_box_embed}
\end{figure}

Since we are interested in the RNN (i.e., LSTM) part of the model, we inject bit errors to the embedding layer and LSTM layer separately and keep the convolutional layer error-free. The prediction accuracy of the LSTM layer and the embedding layer against media errors is explored in Table~\ref{tab:imdb_accu_lstm}, Fig.~\ref{fig:imdb_box_lstm}, Table~\ref{tab:imdb_accu_embed} and Fig.~\ref{fig:imdb_box_embed}. The robustness of the LSTM layer and embedding layer are both $R(0.95)\approx 0.012$. Note that the number of outliers for RNN-based model is more than previous CNN-based image recognition tasks, possibly due to the fact that the weights in LSTM layers are shared by all LSTM cells and a small change of weight will affect all LSTM cells. 

\begin{table}[htbp]
\centering
\caption{Average prediction accuracy for IMDB when the LSTM layer has bit errors}
\label{tab:imdb_accu_lstm}
\begin{tabular}{|c:c:c:c:c:c:c|}
	\hline RBER & 0 & 1e-3 & 2e-3 & 3e-3 & 4e-3 & 5e-3 \\ 
	\hdashline Ave. Accu. & 0.8777 & 0.8715 & 0.8712 & 0.8702 & 0.8650 & 0.8648 \\ 
	\hline RBER & 6e-3 & 7e-3 & 8e-3 & 9e-3 & 1e-2 & 1.1e-2 \\ 
	\hdashline Ave. Accu. & 0.8635 & 0.8596 & 0.8437 & 0.8368 & 0.8373 & 0.8365 \\ 
	\hline RBER & 1.2e-2 & 1.3e-2 & 1.4e-2 & 1.5e-2 & 1.6e-2 & 1.7e-2 \\ 
	\hdashline Ave. Accu. & 0.8366 & 0.8369 & 0.8278 & 0.8281 & 0.8247 & 0.8186 \\ 
	\hline 
\end{tabular} 
\end{table}

\begin{table}[htbp]
	\centering
	\caption{Average prediction accuracy for IMDB when the embedding layer has bit errors}
	\label{tab:imdb_accu_embed}
	\begin{tabular}{|c:c:c:c:c:c:c|}
		\hline RBER & 0 & 1e-3 & 2e-3 & 3e-3 & 4e-3 & 5e-3 \\ 
		\hdashline Ave. Accu. & 0.8777 & 0.8765 & 0.8755 & 0.8709 & 0.8652 & 0.8652 \\ 
		\hline RBER & 6e-3 & 7e-3 & 8e-3 & 9e-3 & 1e-2 & 1.1e-2 \\ 
		\hdashline Ave. Accu. & 0.8548 & 0.8518 & 0.8502 & 0.8446 & 0.8436 & 0.8386 \\ 
		\hline RBER & 1.2e-2 & 1.3e-2 & 1.4e-2 & 1.5e-2 & 1.6e-2 & 1.7e-2 \\ 
		\hdashline Ave. Accu. & 0.8412 & 0.8321 & 0.8231 & 0.8144 & 0.8123 & 0.8130 \\ 
		\hline 
	\end{tabular} 
\end{table}

%
%
%

\section{Weight Nulling by a Check-bit}
In this section, we provide observations and schemes to improve the robustness of neural networks for the same datasets and models in the previous section. The basic approach is to detect (instead of correct) media errors and set the erroneous weights to a default value.

Considering a 16-bit representation of one weight, 15 bits are used to represent the binary expansion of the weights while the last one bit is used as a check-bit such that the modulo-two sum of all 16 bits are zero. In this scenario, if any odd number of bits are flipped from their true values, we can detect that this weight includes erroneous bits. Since each bit error in media occurs with small probability, one-bit error dominates all larger number of errors,. Therefore, it is with high probability that we can detect most of the erroneous weights.

Note that most of the weights are small in their absolute values after training. We set the values of detected erroneous weights as zeros to minimize the distortion on the overall model. This is equivalent to remove the corresponding connection between two neurons and we call it {\em weight nulling}. The rest of the section is devoted to the comparison of robustness of neural network with and without weight nulling.

\subsection{MNIST}
\begin{figure}[htbp]
\centering
\includegraphics[width=1\linewidth]{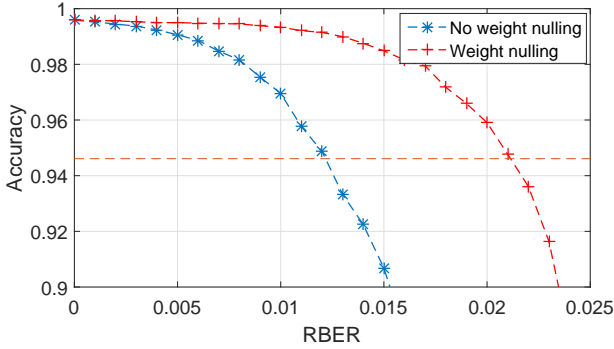}
\caption{Average accuracy of the CNN for MNIST dataset with and without weight nulling.}
\label{fig:mnist_crc}
\end{figure}

Fig.~\ref{fig:mnist_crc} shows the average prediction accuracy of the CNN architecture defined in Table~\ref{tab:mnist} with and without weight nulling technique. The curve labeled as ``No weight nulling'' is a graphical restatement of Table~\ref{tab:mnist_accu}.  The horizontal dashed line indicates $95\%$ accuracy of the undistorted model. It can be observed that the robustness $R(0.95)$ has been increased from $0.012$ to $0.021$ by weight nulling, i.e., to tolerate the same accuracy loss, the RBER of the media can be increased by $75\% $.

\begin{figure}[htbp]
\centering
\includegraphics[width=1\linewidth]{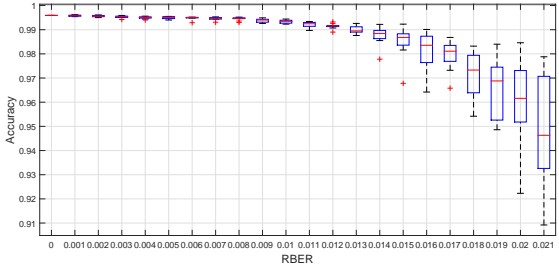}
\caption{Box plot of accuracy for MNIST with weight nulling.}
\label{fig:mnist_crc_naive_boxplot}
\end{figure}

Fig.~\ref{fig:mnist_crc_naive_boxplot} shows the box plot of the prediction accuracy for 40 independently distorted models with weight nulling. Compared to the box plot without weight nulling in Fig.~\ref{fig:mnist_naive_box}, weight nulling not only increases the predication accuracy for the same RBER, it also reduces the variance of accuracy as well.

\subsection{CIFAR-10}
\begin{figure}[htbp]
\centering
\includegraphics[width=1\linewidth]{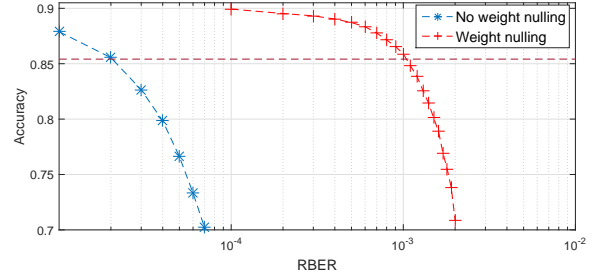}
\caption{Average accuracy of the CNN for CIFAR-10 dataset with and without weight nulling.}
\label{fig:cifar10_crc}
\end{figure}

Fig.~\ref{fig:cifar10_crc} shows the average prediction accuracy of the CNN architecture defined in Table~\ref{tab:cifar10} with and without weight nulling technique. The horizontal dashed line indicates $95\%$ accuracy of the undistorted model. It can be seen that the robustness measure $R(0.95)$ has been increased from $2\times 10^{-5}$ to $10^{-3}$ by weight nulling. That is, the tolerance of RBER for the same accuracy loss has been improved by 50 times.

\begin{figure}[htbp]
\centering
\includegraphics[width=1\linewidth]{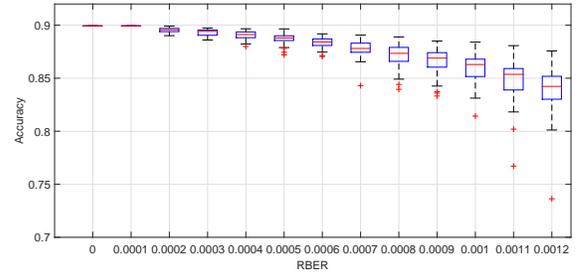}
\caption{Box plot of accuracy for CIFAR-10 with weight nulling.}
\label{fig:cifar10_crc_box}
\end{figure}

Smaller variance of the prediction accuracy from distorted models can also be observed in Fig.~\ref{fig:cifar10_crc_box}.

\subsection{IMDB}
\begin{figure}[htbp]
\centering
\includegraphics[width=1\linewidth]{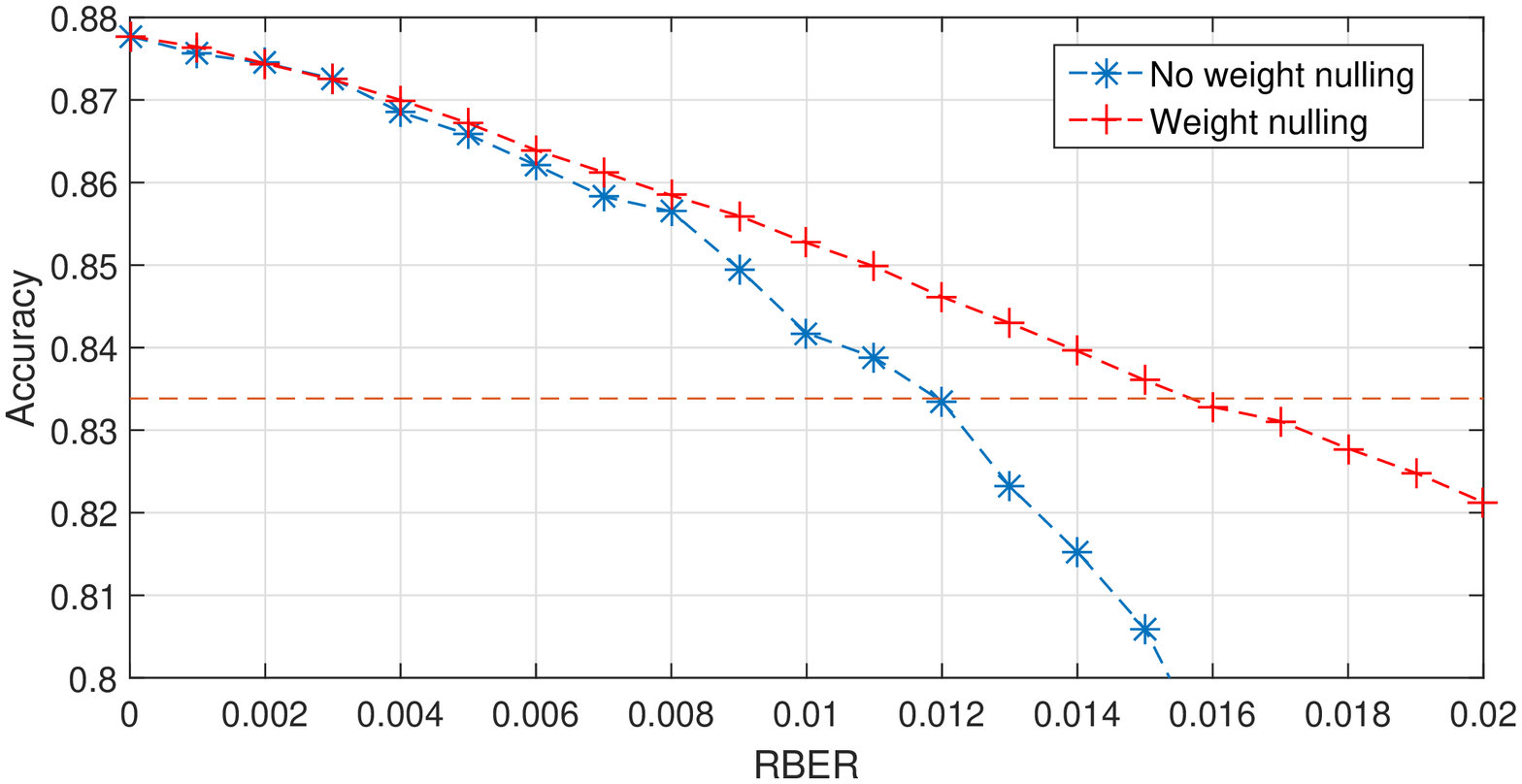}
\caption{Average accuracy of IMDB with and without weight nulling when the LSTM layer has bit errors.}
\label{fig:imdb_crc}
\end{figure}

Fig.~\ref{fig:imdb_crc} shows the average prediction accuracy of the RNN architecture defined in Table~\ref{tab:imdb} with and without weight nulling technique. Bit errors are only applied to either the LSTM layer or embedding layer. The horizontal dashed line indicates $95\%$ accuracy of the undistorted model. It can be seen that the robustness measure $R(0.95)$ for the LSTM and the embedding layer have been increased from $0.012$ to $0.015$ and from $0.012$ to $0.023$, respectively. 

\begin{figure}[htbp]
\centering
\includegraphics[width=1\linewidth]{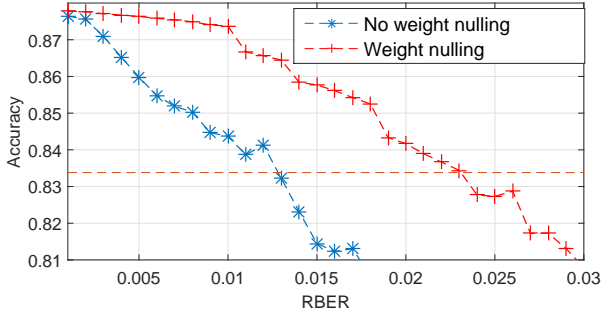}
\caption{Average accuracy of IMDB with and without weight nulling when the embedding layer has bit errors.}
\label{fig:mnist_crc_embed}
\end{figure}

\begin{figure}[htbp]
\centering
\includegraphics[width=1\linewidth]{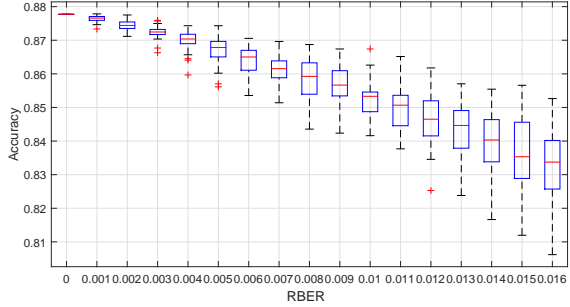}
\caption{Box plot of accuracy for IMDB with weight nulling when the LSTM layer has bit errors.}
\label{fig:imdb_crc_box}
\end{figure}

\begin{figure}[h]
	\centering
\includegraphics[width=1\linewidth]{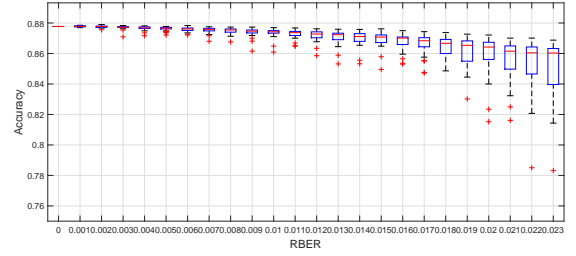}
\caption{Box plot of accuracy for IMDB with weight nulling when the embedding layer has bit errors.}
\label{fig:imdb_crc_embed_box}
\end{figure}

The improvement of average prediction accuracy by weight nulling for RNN-based IMDB classification  is not as large as the other architectures and datasets. This is probably due to the recursive structure of RNN layers where noise on weights has more effect on the overall performance. However, the box plots in Fig.~\ref{fig:imdb_crc_box} and Fig.~\ref{fig:imdb_crc_embed_box} show that the number of outliers is largely reduced compared to Fig.~\ref{fig:imdb_box_lstm} and Fig.~\ref{fig:imdb_box_embed} without weight nulling, which means desirable smaller variance against the noise can also be realized.

\subsection{A summary of weight nulling}
Table~\ref{tab:compare} summarizes the comparison of robustness measure $R(0.95)$ of different datasets and  neural network architectures. We conjecture that for more complicated image recognition tasks with large neural nets, e.g., vgg-16 nets, weight nulling could improve the robustness by orders of magnitude. In fact, we use transfer learning on ImageNet for a similar vgg-16 nets to classify cats and dogs with input image size being 224 by 224. The robust has been improve from $10^{-6}$ to $2.5\times 10^{-4}$, a 250 times more tolerable RBER achieved. For simple tasks where the robustness is already in the orders of $10^{-2}$, the improvement exists but is less significant.
\begin{table}[htbp]
	\centering
	\caption{Robustness improvement of weight nulling}
	\label{tab:compare}
\begin{tabular}{|c|c|c|c|}
	\hline  & MNIST & CIFAR-10 & Cats-Dogs \\ 
	\hline NN architecture & CNN & CNN & CNN \\ 
	\hline  No weight nulling & $0.012$ & $2E{-5}$ & $1E{-6}$  \\ 
	\hline Weight nulling & $0.021$ & $1E{-3}$ &  $2.5E{-4}$  \\ 
	\hline 
\end{tabular} 
\vspace{1em}
\begin{tabular}{|c|c|c|}
	\hline  & IMDB (LSTM layer) & IMDB (embedding layer)  \\ 
	\hline NN architecture  & RNN & RNN \\ 
	\hline  No weight nulling & $0.012$  & $0.012$ \\ 
	\hline Weight nulling & $0.015$ & $0.023$ \\ 
	\hline 
\end{tabular} 
\end{table}

\section{Hamming-distance-based Binary Representation of Numbers}\label{sec:HDB}
Consider the $q$-bit binary expansion representation of real numbers in $[w_{\textrm{min}}, w_{\textrm{max}} ]$. As stated in Section~\ref{sec:bin_exp}, if the leftmost bit (most significant bit) is flipped, then the difference between the original number and the decoded number of the 1-bit erroneous binary representation will be as large as half of the range those $q$ bits can represent. We can say that the relative distortion can be $0.5$, i.e., 
\[
\frac{|w_{\textrm{erroneous}}-w_{\textrm{true}}|}{w_{\textrm{max}} - w_{\textrm{min}}} = \frac{1}{2},
\]
which is independent of $q$. Similar problems occur in IEEE 754 standards if an exponent bits or the sign bit is flipped.

In this section, we provide some theoretical definitions and analysis used for binary representation of real numbers under bit errors. We propose a binary representation, called {\em Hamming-distance-based representation}, of real numbers beyond IEEE 754 standards and the binary expansion. A remarkable feature of the Hamming-distance-based representation is that the relative distortion of the real number is vanishing as the number of bits $n$ for representation increases, conditioned on that a constant number of bits in the binary representation is erroneous. 

\begin{definition}
{\em	
Let $q$ be the number of bits to represent a real number (e.g., weight). Let $\cR=[w_{\textrm{min}},w_{\textrm{max}}]$ be the interval that covers all real numbers needed to represent. Let $\cF = \{0,1,\ldots,2^q-1\}$ be the set of indices after a uniform quantization of $\cR$. Then each real number corresponds to an index in $\cF$. For binary storage media,  $\cF$ is represented by $q$ bits by a bijection $f:\{0,1\}^q \leftrightarrow \cF$. 
We define the distortion under the bijection $f$ between two binary arrays $d(f, b_1, b_2)$ as the normalized $L_1$ distance of their corresponding indices, i.e., 
$$d(f, b_1, b_2) = \frac{|f(b_1)-f(b_2)|}{2^q}, b_1,b_2\in\{0,1\}^q.$$ 
Let $d_H(x_1,x_2)$ be the Hamming distance between two binary arrays $x_1$ and $x_2$. 
We define the maximum and average distance-1 distortion under $f$ of one binary array $b$ as the maximum and average distortion between $b$ and its neighbors whose Hamming distance is 1 from $b$, i.e., 
$$d_{\textrm{max},1}(f,b) = \max_{\{x:d_H(b,x)=1\}} d(f,b,x)$$ 
and 
$$d_{\textrm{ave},1}(f,b) = \frac{\sum_{\{x:d_H(b,x)=1\}} d(f,b,x)}{|\{x:d_H(b,x)=1\}|}.$$
 We then define the distance-1 distortion under the bijection $f$ as 
 $$d_{\textrm{max},1}(f) = \max_{b\in\{0,1\}^q} d_{\textrm{max},1}(f,b)$$
  and 
$$d_{\textrm{ave},1}(f) = \frac{1}{2^q}\sum_{b\in\{0,1\}^q} d_{\textrm{ave},1}(f,b).$$
For a constant number $k$, the definitions of distortion measures $d_{\textrm{max},k}(f,b),d_{\textrm{ave},k}(f,b),d_{\textrm{max},k}(f),$ and $ d_{\textrm{ave},k}(f)$ can be generalized by exploring the distortion of neighbors with Hamming distance less than or equal to $k$. 
 }
\end{definition}

\begin{exam}
	Let $q=2$ and $\cF={0,1,2,3}$. Since the distortion is normalized, the set of real numbers to represent can be any scaled values of the set $\{0,\frac{1}{3},\frac{2}{3},1\}$. If the bijection $f$ is the binary expansion that maps $\{(00),(01),(10),(11)\}\leftrightarrow \cF$ in the corresponding order. Then we have the following example distortion.	
	\begin{align*}
	d_{\textrm{max},1}(f,(00)) &= \max\left\{d(f,(00),(01)),d(f,(00),(10))\right\} \\
		&= \max\left\{\frac{|0-1|}{2^q},\frac{|0-2|}{2^q}  \right\}= \frac{1}{2}.
	\end{align*}
	So $d_{\textrm{max},1}(f) \geq d_{\textrm{max},1}(f,(00)) =  \frac{1}{2}$ and it can be shown that $d_{\textrm{max},1}(f)=  \frac{1}{2}$.
	
	If the bijection $f$ is a length-two Gray code that maps $\{(00),(01),(11),(10)\}\leftrightarrow \cF$ in the corresponding order. Then
	\begin{align*}
	d_{\textrm{max},1}(f,(00)) &= \max\left\{d(f,(00),(01)),d(f,(00),(10))\right\} \\
	&= \max\left\{\frac{|0-1|}{2^q},\frac{|0-3|}{2^q}  \right\}= \frac{3}{4},
	\end{align*}
	thus  $d_{\textrm{max},1}(f)\geq  \frac{3}{4}$. This means that the Gray code mapping is worse than the binary expansion under the defined distortion measure. Also note that as $q$ increases, $d_{\textrm{max},1}(f)$ remains $\frac{1}{2}$ for binary expansion and $d_{\textrm{max},1}(f)$ converges to $1$ for Gray codes.
\end{exam}

\begin{remark}
The range of $\cF$ is actually $2^q-1$ but the denominator is chosen to be $2^q$ to simplify notations since 1) it is proportional with a scale factor $\frac{2^q-1}{2^q}$; 2) when $q$ is large, the scale factor is approximately $1$.
\end{remark}
\begin{remark}
		Gray codes $\cG:\{0,1\}^q \leftrightarrow \{0,1,\ldots,2^q-1\}$ that maps all binary arrays to an integer set satisfy that if $i$ and $j$ are adjacent integers, then their Gray code representations have Hamming distance 1, which is the optimally smallest number. The goal of this section is to propose a ``reverse Gray''-like mapping, where the adjacent binary arrays with Hamming distance 1 should have small distance in their integer representations.
\end{remark}

\begin{const}{
\em
We construct the Hamming-distance-based bijection $f$ as follows.

Let $b_1,b_2\in\{0,1\}^q$, define $b_1 \succeq b_2$ if either one of the following two conditions are satisfied. 1) $w_H(b_1)>w_H(b_2)$, 2) $w_H(b_1)=w_H(b_2)$ and $b_1$ is lexicographically greater than or equal to $b_2$, where $w_H(x)$ is the Hamming weight of a binary array $x$. 

The Hamming-distance-based bijection is defined such that  $\forall b_1, b_2 \in\{0,1\}^q$, $i_1\geq i_2 \Leftrightarrow b_1 \succeq b_2$, where $i_1 = f(b_1)$ and $i_2 = f(b_2)$.  
}
\end{const}

\begin{exam}
Let $q=3$ and $\cF=\{0,1,\ldots,7\}$. The Hamming-distance-based bijection can be defined as $$\{(000),(001),(010),(100),(011),(101),(110),(111)\}\leftrightarrow \cF$$ in the corresponding order. That is, binary arrays with lower Hamming weights are mapped to smaller numbers in $\cF$.

$d_{\textrm{max},1}(f)=\frac{1}{2}$ can be achieved by multiple distortion mechanism. One could be the bit error on the first bit of $(001)$ (representing $1\in\cF$), which ends up to $(101)$ (representing $5\in\cF$). The distortion is then $\frac{5-1}{2^3}=\frac{1}{2}$. For $q=3$, the Hamming-distance-based bijection has the same $d_{\textrm{max},1}(f)$ as the binary expansion. But for larger $q$, the distortion of Hamming-distance-based bijection would decrease to $0$, while the binary expansion would remain $\frac{1}{2}$.

\end{exam}

\begin{theorem}\label{thm:hdb}{
		\em	
	Let $f$ be the Hamming-distance-based bijection and $k$ be a constant number, then
	\[
	\lim_{q\rightarrow \infty} d_{\textrm{max},k}(f) = d_{\textrm{ave},k}(f) = 0.
	\]
	
}
\end{theorem}
\begin{IEEEproof}
According to the definition of $d_{\textrm{max},k}(f)$ and $d_{\textrm{ave},k}(f)$, we have the following equations
\begin{align*}
d_{\textrm{ave},k}(f) \leq d_{\textrm{max},k}(f) \leq kd_{\textrm{max},1}(f),
\end{align*}
where the second inequality is due to that a  binary array can be converted to another binary array with Hamming distance $k$ by $k$ steps where one bit is changed in each step. Therefore, we only need to prove  
$$\lim_{q\rightarrow \infty}d_{\textrm{max},1}(f)= 0.$$

By the construction of Hamming-distance-based bijection, the first binary array is the all-zero array (i.e., the Hamming weight is $0$). The next $q$ binary arrays all have Hamming weight $1$ and so on. Thus we can sequentially partition all $2^q$ binary arrays into $q+1$ groups, where each group are all binary arrays with Hamming weight $k$ and the size of each group is $q \choose k$ for $k=0,1,\ldots,q$. If any bit in a binary array is flipped, then it is converted to another binary array in one of the adjacent groups. The distortion (their corresponding decimal representation in $\cF$) is less than or equal to the size of the two groups, i.e.,
\[
d_{\textrm{max},1}(f,b) \leq \frac{ {q \choose k} + {q \choose k+1} }{2^q},
\]
if $w_H(b)<\frac{q}{2}$; otherwise,
\[
d_{\textrm{max},1}(f,b) \leq \frac{{q \choose k} + {q \choose k-1}}{2^q}.
\]
Thus,
\begin{align*}
\lim_{q\rightarrow\infty} d_{\textrm{max},1}(f) &= \lim_{q\rightarrow\infty} \max_{b\in\{0,1\}^q} d_{\textrm{max},1}(f,b) \\
& \leq \lim_{q\rightarrow\infty} \frac{ 2{q \choose q/2} }{2^q} \\
& = \lim_{q\rightarrow\infty} \frac{2\sqrt{\frac{2}{\pi q}} 2^q}{2^q}\\
& = \lim_{q\rightarrow\infty} 2\sqrt{\frac{2}{\pi q}} \\
& = 0,
\end{align*}
where the first equality follows the definition, the second inequality follows that $q \choose k$ is maximized at $k=\frac{q}{2}$, and the third equality follows from the Stirling approximation.

\end{IEEEproof}

\begin{remark}
According Theorem~\ref{thm:hdb}, the distortion decreases to 0 as $q$ increases, which is a much more promising result than binary expansion and IEEE 754 floating-point number representation where the distortion is at least a constant. But it can also be observed that $d_{\textrm{max},1}(f)$ decreases very slowly in the order of $\frac{1}{\sqrt q}$.
\end{remark}

\begin{figure}[htbp]
\centering
\includegraphics[width=1\linewidth]{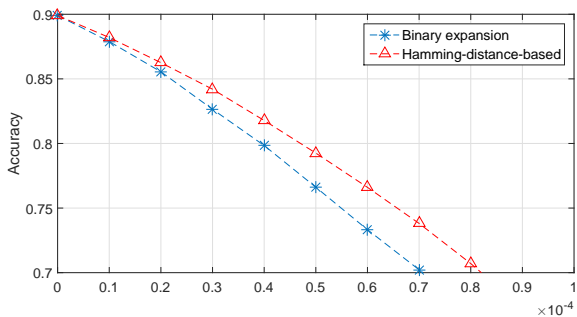}
\caption{Accuracy of CIFAR-10 for binary expansion and Hamming-distance-based representation.}
\label{fig:cifar10_wc}
\end{figure}

Fig.~\ref{fig:cifar10_wc} shows the comparison of CIFAR-10 dataset average prediction accuracy for binary expansion and Hamming-distance-based representation when media errors exist. The curve corresponding to ``binary expansion'' is a restatement of Table~\ref{tab:cifar10_accu} and the curve labeled ``No weight nulling'' in Fig.~\ref{fig:cifar10_crc}. It can be observed that the accuracy is slightly improved using the Hamming-distance-based bijection. We conjecture that the limited improvement is due to the fact that $q=16$ is not sufficiently large in the experiment.

\section{Conclusions}
In this paper, we study the robustness of neural networks against media errors for different neural network architectures and datasets. We propose a weight nulling algorithm based on a check bit to improve the robustness. We also provide a Hamming-distance-based binary representation of real numbers such that the distortion brought by bit errors in stored media can be reduced compared to a simple binary expansion representation or IEEE Standard for Floating-Point Arithmetic (IEEE 754).

\bibliographystyle{IEEEtran}
\bibliography{reference_minghai}

\begin{thebibliography}{10}
\providecommand{\url}[1]{#1}
\csname url@samestyle\endcsname
\providecommand{\newblock}{\relax}
\providecommand{\bibinfo}[2]{#2}
\providecommand{\BIBentrySTDinterwordspacing}{\spaceskip=0pt\relax}
\providecommand{\BIBentryALTinterwordstretchfactor}{4}
\providecommand{\BIBentryALTinterwordspacing}{\spaceskip=\fontdimen2\font plus
\BIBentryALTinterwordstretchfactor\fontdimen3\font minus
  \fontdimen4\font\relax}
\providecommand{\BIBforeignlanguage}[2]{{%
\expandafter\ifx\csname l@#1\endcsname\relax
\typeout{** WARNING: IEEEtran.bst: No hyphenation pattern has been}%
\typeout{** loaded for the language `#1'. Using the pattern for}%
\typeout{** the default language instead.}%
\else
\language=\csname l@#1\endcsname
\fi
#2}}
\providecommand{\BIBdecl}{\relax}
\BIBdecl

\bibitem{Schmidhuber15}
J.~Schmidhuber, ``Deep learning in neural networks: An overview,'' \emph{Neural
  Networks}, vol.~61, pp. 85--117, 2015, published online 2014; based on TR
  arXiv:1404.7828 [cs.NE].

\bibitem{HTF01}
T.~Hastie, R.~Tibshirani, and J.~Friedman, \emph{The Elements of Statistical
  Learning}.\hskip 1em plus 0.5em minus 0.4em\relax Springer, New York, 2001.

\bibitem{JKR09}
K.~Jarrett, K.~Kavukcuoglu, and M.~Ranzato, ``What is the best multi-stage
  architecture for object recognition?'' in \emph{IEEE Int. Conf. Computer
  Vision}, Kyoto, Japan, September 2009, p. 2146–2153.

\bibitem{KSH12}
\BIBentryALTinterwordspacing
A.~Krizhevsky, I.~Sutskever, and G.~Hinton, ``Imagenet classification with deep
  convolutional neural networks,'' in \emph{Advances in Neural Information
  Processing Systems 25}, F.~Pereira, C.~J.~C. Burges, L.~Bottou, and K.~Q.
  Weinberger, Eds.\hskip 1em plus 0.5em minus 0.4em\relax Curran Associates,
  Inc., 2012, pp. 1097--1105. [Online]. Available:
  \url{http://papers.nips.cc/paper/4824-imagenet-classification-with-deep-convolutional-neural-networks.pdf}
\BIBentrySTDinterwordspacing

\bibitem{GLF09}
A.~Graves, M.~Liwicki, S.~Fernandez, R.~Bertolami, H.~Bunke, and
  J.~Schmidhuber, ``A novel connectionist system for improved unconstrained
  handwriting recognition,'' \emph{IEEE Trans. Pattern Analysis and Machine
  Intelligence}, vol.~31, no.~5, pp. 855 -- 868, May 2009.

\bibitem{HAF14}
H.~Sak, A.~Senior, and F.~Beaufays, ``Long short-term memory recurrent neural
  network architectures for large scale acoustic modeling,'' 2014.

\bibitem{GSKSS16}
K.~Greff, R.~Srivastava, J.~Koutník, B.~Steunebrink, and J.~Schmidhuber,
  ``{LSTM}: A search space odyssey,'' \emph{IEEE Trans. Neural Networks and
  Learning Systems}, pp. 1--11, July 2016.

\bibitem{RB93}
M.~Riedmiller and H.~Braun, ``A direct adaptive method for faster
  backpropagation learning: the {RPROP} algorithm,'' in \emph{IEEE Int. Conf.
  Neural Networks}, San Francisco, CA, USA, March 2009, pp. 586--591.

\bibitem{Nielsen89}
R.~Hecht-Nielsen, ``Theory of the backpropagation neural network,'' in
  \emph{Int. Joint Conf. on Neural Networks (IJCNN)}, Washington, DC, USA,
  1989, pp. 593--605.

\bibitem{RL09}
W.~Ryan and S.~Lin, \emph{Channel Codes: Classical and Modern}.\hskip 1em plus
  0.5em minus 0.4em\relax Cambridge University Press, 2009.

\bibitem{CW17}
N.~Carlini and D.~Wagner, ``Towards evaluating the robustness of neural
  networks,'' in \emph{IEEE Symp. on Security and Privacy (SP)}, San Jose, CA,
  USA, 2017, pp. 39--57.

\bibitem{ZSLG16}
S.~Zheng, Y.~Song, T.~Leung, and I.~Goodfellow, ``Improving the robustness of
  deep neural networks via stability training,'' in \emph{IEEE Conf. on
  Computer Vision and Pattern Recognition (CVPR)}, Las Vegas, NV, USA, 2016,
  pp. 4480 -- 4488.

\bibitem{CBD15}
\BIBentryALTinterwordspacing
M.~Courbariaux, Y.~Bengio, and J.~David, ``Binaryconnect: Training deep neural
  networks with binary weights during propagations,'' 2015. [Online].
  Available: \url{http://arxiv.org/abs/1511.00363}
\BIBentrySTDinterwordspacing

\bibitem{MAAEM16}
\BIBentryALTinterwordspacing
P.~Merolla, R.~Appuswamy, J.~Arthur, S.~Esser, and D.~Modha, ``Deep neural
  networks are robust to weight binarization and other non-linear
  distortions,'' 2016. [Online]. Available:
  \url{http://arxiv.org/abs/1606.01981}
\BIBentrySTDinterwordspacing

\bibitem{CT91}
T.~Cover and J.~Thomas, \emph{Elements of Information Theory}.\hskip 1em plus
  0.5em minus 0.4em\relax Wiley-Interscience New York, NY, USA, 1991.

\bibitem{SZ14}
\BIBentryALTinterwordspacing
K.~Simonyan and A.~Zisserman, ``Very deep convolutional networks for
  large-scale image recognition,'' in \emph{Proc. International Conference on
  Learning Representations}, 2014. [Online]. Available:
  \url{http://arxiv.org/abs/1409.1556 (2014)}
\BIBentrySTDinterwordspacing

\bibitem{MSCCD13}
\BIBentryALTinterwordspacing
T.~Mikolov, I.~Sutskever, K.~Chen, G.~Corrado, and J.~Dean, ``Distributed
  representations of words and phrases and their compositionality,'' 2013.
  [Online]. Available: \url{https://arxiv.org/abs/1310.4546}
\BIBentrySTDinterwordspacing

\end{thebibliography}

\end{document}